# Three-Receiver Broadcast Channels with Side Information

Saeed Hajizadeh
(Undergraduate Student)
Department of Electrical Engineering
Ferdowsi University of Mashhad
Mashhad, Iran
Saeed.hajizadeh1367@gmail.com

Ghosheh Abed Hodtani
Department of Electrical Engineering
Ferdowsi University of Mashhad
Mashhad, Iran
ghodtani@gmail.com

*Abstract*—Three-receiver broadcast channel (BC) is of interest due to its information theoretical differences with two receiver one. In this paper, we derive achievable rate regions for two classes of 3-receiver BC with side information available at the transmitter, Multilevel BC and 3-receiver less noisy BC, by using superposition coding, Gel'fand-Pinsker binning scheme and Nair-El Gamal indirect decoding. Our rate region for multilevel BC subsumes the Steinberg rate region for 2-receiver degraded BC with side information as its special case. We also find the capacity region of 3-receiver less noisy BC when side information is available both at the transmitter and at the receivers.

*Keywords: 3-receiver broadcast channel, less noisy, Multilevel broadcast channel*

## I. Introduction

The k-receiver, $k \geq 3$, broadcast channel (BC) was first studied by Borade *et al.* in [1] where they simply surmised that straightforward extension of Körner-Marton's capacity region for two-receiver BCs with degraded message sets [2] to k-receiver multilevel broadcast networks is optimal. Nair-El Gamal [3] showed that the capacity region of a special class of 3-receiver BCs with two degraded message sets when one of the receivers is a degraded version of the other, is a superset of [1], thus proving that direct extension of [2] is not in general optimal. Nair and Wang later in [4] established the capacity region of the 3-receiver less noisy BC. Channels with Side information (SI), were first studied by Shannon [5], where he found the capacity region of the Single-Input-Single-Output channel when SI is causally available at the encoder. Gelf'and and Pinsker [6] found the capacity region of a single-user channel when SI is non-causally available at the transmitter while the receiver is kept ignorant of it. Cover and Chiang [7] extended the results of [6] to the case where SI is available at both the encoder and the decoder. Multiple user channels with side information were studied in [8] where inner and outer bounds for degraded BC with non-causal SI and capacity region of degraded BC with causal SI were found. Moreover, in [9] inner and outer bounds were given to general two-user BCs with SI available at the transmitter and other special cases both for BCs and MACs were also found.

In this paper, we find the achievable rate region of Multilevel BC and 3-receiver less noisy BC both with SI non-causally available at the encoder. Our achievable rate regions reduce to that of [3] and [4] when there is no side information. We also find the capacity region of the latter when side information is also available at the receivers. The rest of the paper is organized as follows. In section II, basic definitions and notations are presented. In sections III and IV, new achievable rate regions are given for the Multilevel BC and 3-receiver less noisy BC, respectively. In section V, conclusion is given.

## II. Definitions

Random variables and their realizations are denoted by uppercase and lowercase letters, respectively, e.g. x is a realization of X. Let $\mathcal{X}, \mathcal{Y}_1, \mathcal{Y}_2, \mathcal{Y}_3, and\ \mathcal{S}$ be finite sets showing alphabets of random variables. The n-sequence of a random variable is given by $X^n$ where the superscript is omitted when the choice of n is clear, thus we only use boldface letters for the random variable itself, i.e. $\mathbf{x} = x^n$. Throughout, we assume that $X_i^n$ is the sequence $(X_i, X_{i+1}, \dots, X_n)$.

***Definition 1:*** A channel $X \to Z$ is said to be a degraded version of the channel $X \to Y$ with SI if $X \to Y \to Z$ be a Markov chain conditioned on every $s \in \mathcal{S}$ for all $p(u, x|s)$.

Multilevel BC with side information, denoted by $(\mathcal{X}, \mathcal{S},\ \mathcal{Y}_1, \mathcal{Y}_2, \mathcal{Y}_3, p(y_1, y_3|x, s), p(y_2|y_1))$, is a 3-receiver BC with 2-degraded message sets with input alphabet $\mathcal{X}$ and output alphabets $\mathcal{Y}_1, \mathcal{Y}_2$, and $\mathcal{Y}_3$. The side information is the random variable S distributed over the set $\mathcal{S}$ according to $p(s)$. The transition probability function $p(y_1, y_3|x, s)$ describes the relationship between channel input X, side information S, and channel outputs $Y_1$ and $Y_3$ while the probability function $p(y_2|y_1)$ shows the virtual channel modeling the output $Y_2$ as the degraded version of $Y_1$. Independent message sets $m_0 \in \mathcal{M}_0$ and $m_1 \in \mathcal{M}_1$ are to be reliably sent, $m_0$ being the common message for all the receivers and $m_1$ the private message only for $Y_1$. Channel model is depicted in Fig. 1.

***Definition 2:*** A $(n, 2^{nR_0}, 2^{nR_1}, \epsilon)$ two-degraded message set code for the Multilevel BC with side information $\big(p(y_1, y_3|x, s), p(y_2|y_1)\big)$ consists of an encoder map

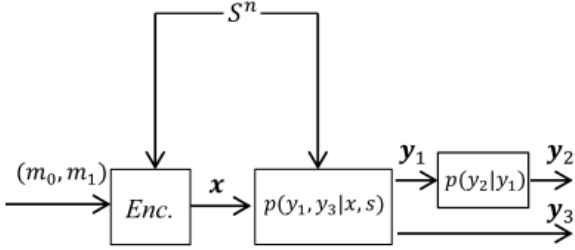

Figure 1. Multilevel broadcast channel with side information.

$$f : \{1,2,\ldots,M_0\} \times \{1,2,\ldots,M_1\} \times \mathcal{S}^n \to \mathcal{X}^n$$

and a tuple of decoding maps

$$g_{y_1} : \mathcal{Y}_1^n \to \{1,2,\ldots,M_0\} \times \{1,2,\ldots,M_1\}$$
$$g_{y_2} : \mathcal{Y}_2^n \to \{1,2,\ldots,M_0\}$$
$$g_{y_3} : \mathcal{Y}_3^n \to \{1,2,\ldots,M_0\}$$

Such that $P_e^{(n)} \leq \epsilon$, i.e.

$$\frac{1}{M_0 M_1} \sum_{m_0=1}^{M_0} \sum_{m_1=1}^{M_1} \sum_{s^n \in \mathcal{S}^n} p(s) p\{g_{y_1}(\mathbf{y}_1) \neq (m_0, m_1) \text{ or }$$
$$g_{y_2}(\mathbf{y}_2) \neq m_0 \text{ or } g_{y_3}(\mathbf{y}_3) \neq m_0 | \mathbf{s}, \mathbf{x}(m_0, m_1, \mathbf{s})\} \leq \epsilon$$

The rate pair of the code is defined as

$$(R_0, R_1) = \frac{1}{n}(\log M_0, \log M_1)$$

A rate pair $(R_0, R_1)$ is said to be $\epsilon$-achievable if for any $\eta > 0$ there is an integer $n_0$ such that for all $n \geq n_0$ we have a $(n, 2^{n(R_0-\eta)}, 2^{n(R_1-\eta)}, \epsilon)$ code for $(p(y_1, y_3|x,s), p(y_2|y_1))$. The union of the closure of all $\epsilon$-achievable rate pairs is called the capacity region $\mathcal{C}_{MBC}$.

***Definition 3:*** A channel $X \to Y$ is said to be less noisy than the channel $X \to Z$ in the presence of side information if

$$I(U;Y|S=s) \geq I(U;Z|S=s)$$
$$\forall p(u,x,y,z|s) = p(u|s)p(x|u,s)p(y,z|x,s) \text{ and } \forall s \in \mathcal{S}.$$

The 3-receiver less noisy BC with side information is depicted in Fig. 2, where $Y_1$ is less noisy than $Y_2$ and $Y_2$ is less noisy than $Y_3$, i.e. according to [4], $Y_1 \succcurlyeq Y_2 \succcurlyeq Y_3$.

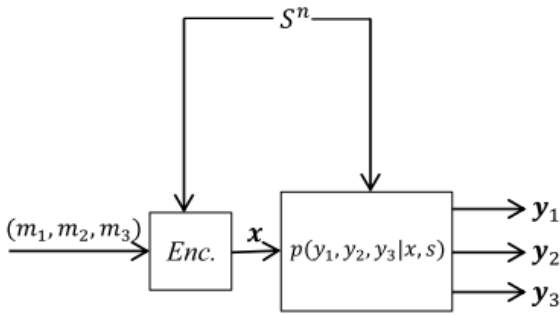

Figure.2. Three-receiver less noisy broadcast channel with side information.

The messages $m_1 \in \mathcal{M}_1$, $m_2 \in \mathcal{M}_2$, $m_3 \in \mathcal{M}_3$ are to be reliably sent to receivers $Y_1, Y_2$, and $Y_3$, respectively. The code and rate tuple definitions are as follows

$$(n, 2^{nR_1}, 2^{nR_2}, 2^{nR_3}, \epsilon)$$
$$(R_1, R_2, R_3) = \frac{1}{n}(\log M_1, \log M_2, \log M_3)$$

Achievable rate tuples and the achievable rate region and the capacity region $\mathcal{C}_L$ are defined in just the same way as Multilevel BC.

## III. MULTILEVEL BROADCAST CHANNEL WITH SIDE INFORMATION

Define $\mathcal{P}$ as the collection of all random variables $(U, V, S, X, Y_1, Y_2, Y_3)$ with finite alphabets such that

$$p(u,v,s,x,y_1,y_2,y_3) =$$
$$p(s)p(u|s)p(v|u,s)p(x|v,s)p(y_1,y_3|x,s)p(y_2|y_1) \quad (1)$$

By (1), the following Markov chains hold:

$$(U,V) \to (X,S) \to (Y_1, Y_3) \quad (2)$$
$$(S, X, Y_3) \to Y_1 \to Y_2 \quad (3)$$

***Theorem 1:*** A pair of nonnegative numbers $(R_0, R_1)$ is achievable for Multilevel BC with side information non-causally available at the transmitter provided that

$$R_0 \leq \min\{I(U;Y_2) - I(U;S), I(V;Y_3) - I(UV;S)\}$$
$$R_1 \leq I(X;Y_1|U) - I(V;S|U) - I(X;S|V) \quad (4)$$
$$R_0 + R_1 \leq I(V;Y_3) + I(X;Y_1|V) - I(X;S|V) - I(UV;S)$$

for some $(U, V, S, X, Y_1, Y_2, Y_3) \in \mathcal{P}$.

***Corollary 1.1:*** By setting $S \equiv \emptyset$ in (4), our achievable rate region in Theorem 1 is reduced to the capacity region of Multilevel BC given in [3].

***Corollary 1.2:*** By setting $Y_3 = Y_1$ and $V = U$ in (4), our achievable rate region reduces to that of [8] for the two-user degraded BC with side information.

***Proof:*** Fix n and a joint distribution on $\mathcal{P}$. Note that side information is distributed i.i.d according to

$$p(\mathbf{s}) = \prod_{i=1}^{n} p(s_i)$$

Split the $\mathcal{M}_1$ message into two independent submessage sets $\mathcal{M}_{11}$, and $\mathcal{M}_{12}$ so that $R_1 = R_{11} + R_{12}$.

***Codebook Generation:*** First randomly and independently generate $2^{n(R_0' + R_0)}$ sequences $\mathbf{u}(m_0', m_0)$, $m_0' \in \{1,2,\ldots,2^{nR_0'}\}$, $m_0 \in \{1,2,\ldots,2^{nR_0}\}$, each one i.i.d according to $\prod_{i=1}^{n} p(u_i)$ and then randomly throw them into $2^{nR_0}$ bins. It is clear that we have $2^{nR_0'}$ sequences in each bin.

Now for each $\mathbf{u}(m_0', m_0)$, randomly and independently generate $2^{n(R_{11}' + R_{11})}$ sequences $\mathbf{v}(m_0', m_0, m_{11}', m_{11})$, $m_{11}' \in$

$\{1, ..., 2^{nR'_{11}}\}$, $m_{11} \in \{1, ..., 2^{nR_{11}}\}$ each one i.i.d according to $\prod_{i=1}^{n} p_{V|U}(v_i|u_i(m'_0, m_0))$, and randomly throw them into $2^{nR_{11}}$ bins.

Now for each sequence $\boldsymbol{v}(m'_0, m_0, m'_{11}, m_{11})$, randomly and independently generate $2^{n(R'_{12}+R_{12})}$ sequences $\boldsymbol{x}(m'_0, m_0, m'_{11}, m_{11}, m'_{12}, m_{12})$ each one i.i.d according to $\prod_{i=1}^{n} p_{X|U,V}(x_i|v_i, u_i) = \prod_{i=1}^{n} p_{X|V}(x_i|v_i)$. Then randomly throw them into $2^{nR_{12}}$ bins. Then provide the transmitter and all the receivers with bins and their codewords.

***Encoding:*** We are given the side information $\boldsymbol{s}$ and the message pair $(m_0, m_1)$. Indeed, our messages are bin indices. We find $m_{11}$, and $m_{12}$. Now in the bin $m_0$ of $\boldsymbol{u}$ sequences look for a $m'_0$ such that $(\boldsymbol{u}(m'_0, m_0), \boldsymbol{s}) \in A_\epsilon^{(n)}$, i.e. the sequence $\boldsymbol{u}$ that is jointly typical with the $\boldsymbol{s}$ given where definitions of typical sequences are given in [12]. Then in the bin $m_{11}$ of $\boldsymbol{v}$ sequences look for some $m'_{11}$ such that

$(\boldsymbol{u}(m'_0, m_0), \boldsymbol{v}(m'_0, m_0, m'_{11}, m_{11}), \boldsymbol{s}) \in A_\epsilon^{(n)}$

Now in the bin $m_{12}$ of $\boldsymbol{x}$ sequences look for some $m'_{12}$ such that

$(\boldsymbol{u}(m'_0, m_0), \boldsymbol{v}(m'_0, m_0, m'_{11}, m_{11}),$
$\boldsymbol{x}(m'_0, m_0, m'_{11}, m_{11}, m'_{12}, m_{12}), \boldsymbol{s}) \in A_\epsilon^{(n)}$

We send the found $\boldsymbol{x}$ sequence. Before bumping into decoding, assume that the correct indices are found through the encoding procedure, i.e. $m'_0 = M'_0, m'_{11} = M'_{11}$ and $m'_{12} = M'_{12}$.

***Decoding:*** Since the messages are uniformly distributed over their respective ranges, we can assume, without loss of generality, that the tuple $(m_0, m_{11}, m_{12}) = (1,1,1)$ is sent.

The second receiver $Y_2$ receives $\boldsymbol{y}_2$ thus having the following error events

$E_{21} = \{(\boldsymbol{u}(M'_0, 1), \boldsymbol{y}_2) \notin A_\epsilon^{(n)}\}$
$E_{22} = \{(\boldsymbol{u}(m'_0, m_0), \boldsymbol{y}_2) \in A_\epsilon^{(n)} \text{ for some } m_0 \neq 1$
  $\text{and } m'_0 \neq M'_0\}$

***Remark 1:*** The following error event

$E_{23} = \{(\boldsymbol{u}(M'_0, m_0), \boldsymbol{y}_1) \in A_\epsilon^{(n)} \text{ for some } m_0 \neq 1\}$

leads us to a redundant inequality.

Now by the weak law of large numbers (WLLN) [14], $p(E_{21}) \leq \epsilon, \forall \epsilon > 0$ as $n \to \infty$. For the second error event we have

$p(E_{22}) = \sum_{m_0, m'_0} \sum_{A_\epsilon^{(n)}} p(\boldsymbol{u}) p(\boldsymbol{y}_2) \leq 2^{n(R_0+R'_0)} 2^{n(H(U,Y_2)+\epsilon)}$
$2^{-n(H(U)-\epsilon)} 2^{-n(H(Y_2)-\epsilon)} = 2^{-n(I(U;Y_2)-3\epsilon-R_0-R'_0)}$

We see that $\forall \epsilon > 0$, $p(E_{22}) \leq \epsilon$ as $n \to \infty$ provided that

$R_0 + R'_0 \leq I(U; Y_2) - 3\epsilon$ (5)

The first receiver $Y_1$ receives $\boldsymbol{y}_1$ and needs to decode both $m_0$ and $m_1$. Therefore, the error events are

$E_{11} = \{(\boldsymbol{u}(M'_0, 1), \boldsymbol{v}(M'_0, 1, M'_{11}, 1),$
  $\boldsymbol{x}(M'_0, 1, M'_{11}, 1, M'_{12}, 1), \boldsymbol{y}_1) \notin A_\epsilon^{(n)}\}$
$E_{12} = \{(\boldsymbol{u}(M'_0, 1), \boldsymbol{v}(M'_0, 1, M'_{11}, 1),$
  $\boldsymbol{x}(M'_0, 1, M'_{11}, 1, m'_{12}, m_{12}), \boldsymbol{y}_1) \in A_\epsilon^{(n)}$
  $\text{for some } m_{12} \neq 1 \text{ and } m'_{12} \neq M'_{12}\}$
$E_{13} = \{(\boldsymbol{u}(M'_0, 1), \boldsymbol{v}(M'_0, 1, m'_{11}, m_{11}),$
  $\boldsymbol{x}(M'_0, 1, m'_{11}, m_{11}, m'_{12}, m_{12}), \boldsymbol{y}_1) \in A_\epsilon^{(n)}$
  $\text{for some } m_{1i} \neq 1 \text{ and } m'_{1i} \neq M'_{1i}, i = 1,2\}$
$E_{14} = \{(\boldsymbol{u}(m'_0, m_0), \boldsymbol{v}(m'_0, m_0, m'_{11}, m_{11}),$
  $\boldsymbol{x}(m'_0, m_0, m'_{11}, m_{11}, m'_{12}, m_{12}), \boldsymbol{y}_1) \in A_\epsilon^{(n)}$
  $\text{for some } m_0 \neq 1 \text{ and } m'_0 \neq M'_0 \text{ and some}$
  $m_{1i} \neq 1 \text{ and } m'_{1i} \neq M'_{1i}, i = 1,2\}$

The first receiver's probability of error can be arbitrarily made small provided that

$R_{12} + R'_{12} \leq I(X; Y_1|V) - 6\epsilon$ (6)
$R_{11} + R'_{11} + R_{12} + R'_{12} \leq I(X; Y_1|U) - 6\epsilon$ (7)
$R_0 + R'_0 + R_{11} + R'_{11} + R_{12} + R'_{12} \leq I(X; Y_1) - 5\epsilon$ (8)

The third receiver $Y_3$ receives $\boldsymbol{y}_3$ and needs to decode only the common message indirectly by decoding the message $m_{11}$. The error events are

$E_{31} = \{(\boldsymbol{u}(M'_0, 1), \boldsymbol{v}(M'_0, 1, M'_{11}, 1), \boldsymbol{y}_3) \notin A_\epsilon^{(n)}\}$
$E_{32} = \{(\boldsymbol{u}(M'_0, 1), \boldsymbol{v}(M'_0, 1, m'_{11}, m_{11}), \boldsymbol{y}_3) \in A_\epsilon^{(n)} \text{ for}$
  $\text{some } m'_{11} \neq 1 \text{ and } m'_{11} \neq M'_{11}\}$
$E_{33} = \{(\boldsymbol{u}(m'_0, m_0), \boldsymbol{v}(m'_0, m_0, m'_{11}, m_{11}), \boldsymbol{y}_3) \in A_\epsilon^{(n)} \text{ for}$
  $\text{some } m_0 \neq 1, m_{11} \neq 1, m'_0 \neq M'_0, \text{and } m'_{11} \neq M'_{11}\}$

Again by using WLLN and AEP, we see that the third receiver's error probabilities can be arbitrarily made small as $n \to \infty$ provided that

$R_0 + R'_0 + R_{11} + R'_{11} \leq I(V; Y_3) - 3\epsilon$ (9)

Using Gel'fand-Pinsker coding we see that the encoders can choose the proper $m'_0, m'_{11}, \text{and } m'_{12}$ indices with vanishing probability of error provided that for every $\epsilon > 0$ and sufficiently large n

$R'_0 \geq I(U; S) + 2\epsilon$ (10)
$R'_{11} \geq I(V; S|U) + 2\epsilon$ (11)
$R'_{12} \geq I(X; S|V) + 2\epsilon$ (12)

Now combining (5) - (9) and (10) - (12) and noting that

$I(V; S|U) + I(U; S) = I(VU; S)$ (13)

and using Fourier-Motzkin procedure afterwards to eliminate $R_{11}$ and $R_{12}$, we obtain (4) as an achievable rate region for Multilevel BC with side information. ∎

## IV. THREE-RECEIVER LESS NOISY BROADCAST CHANNEL WITH SIDE INFORMATION

Define $\mathcal{P}^*$ as the collection of all random variables $(U, V, S, X, Y_1, Y_2, Y_3)$ with finite alphabets such that

$$p(u, v, s, x, y_1, y_2, y_3) = p(s)p(u|s)p(v|u,s)p(x|v,s)p(y_1, y_2, y_3|x, s) \quad (14)$$

**Theorem 2:** A rate triple $(R_1, R_2, R_3)$ is achievable for 3-receiver less noisy BC with side information non-causally available at the transmitter provided that

$$R_1 \leq I(X; Y_1|V) - I(X; S|V)$$
$$R_2 \leq I(V; Y_2|U) - I(V; S|U) \quad (15)$$
$$R_3 \leq I(U; Y_3) - I(U; S)$$

for some joint distribution on $\mathcal{P}^*$.

**Corollary 2.1:** By setting $S \equiv \emptyset$ in the above rate region, it reduces to the capacity region of 3-receiver less noisy BC given in [4].

*Proof:* The proof uses Cover's superposition [15] and Gel'fand-Pinsker random binning coding [6] procedures along with Nair's indirect decoding and is similar to the last proof provided and thus only an outline is provided.

Fix n and a distribution on $\mathcal{P}^*$.

Again note that side information is distributed i.i.d according to

$$p(s) = \prod_{i=1}^{n} p(s_i)$$

Randomly and independently generate $2^{n(R_3' + R_3)}$ sequences $\mathbf{u}(m_3', m_3)$, each distributed i.i.d according to $\prod_{i=1}^{n} p(u_i)$ and randomly throw them into $2^{nR_3}$ bins.

For each $\mathbf{u}(m_3', m_3)$, randomly and independently generate $2^{n(R_2' + R_2)}$ sequences $\mathbf{v}(m_3', m_3, m_2', m_2)$ each distributed i.i.d according to $\prod_{i=1}^{n} p_{V|U}(v_i|u_i)$ and randomly throw them into $2^{nR_2}$ bins.

Now for each generated $\mathbf{v}(m_3', m_3, m_2', m_2)$, randomly and independently generate $2^{n(R_1' + R_1)}$ sequences $\mathbf{x}(m_3', m_3, m_2', m_2, m_1', m_1)$, each one distributed i.i.d according to $\prod_{i=1}^{n} p_{X|V}(x_i|v_i)$ and randomly throw them into $2^{nR_1}$ bins.

Encoding is succeeded with small probability of error provided that

$$R_3' \geq I(U; S) \quad (16)$$
$$R_2' \geq I(V; S|U) \quad (17)$$
$$R_1' \geq I(X; S|V) \quad (18)$$

and decoding is succeeded if

$$R_3 + R_3' \leq I(U; Y_3) \quad (19)$$
$$R_2 + R_2' \leq I(V; Y_2|U) \quad (20)$$
$$R_1 + R_1' \leq I(X; Y_1|V) \quad (21)$$

Now combining (16), (17) and (18) with (19), (20) and (21) gives us (15). ∎

**Theorem 3:** The capacity region of the 3-receiver less noisy BC with side information, non-causally available at the transmitter and the receivers is the set of all rate triples $(R_1, R_2, R_3)$ such that

$$R_1 \leq I(X; Y_1|VS)$$
$$R_2 \leq I(V; Y_2|US) \quad (22)$$
$$R_3 \leq I(U; Y_3|S)$$

**Proof:**

**Achievability:** The direct part of the proof is achieved if you set $\widetilde{Y}_k = (Y_k, S), k = 1,2,3$ in (15).

**Converse:** The converse part uses an extension of lemma 1 in [4].

**Lemma 1:** [4] Let the channel $X \to Y$ be less noisy than the channel $X \to Z$. Consider $(M, S^n)$ to be any random vector such that

$$(M, S^n) \to X^n \to (Y^n, Z^n)$$

forms a Markov chain. Then

1. $I(Y^{i-1}; Z_i|M, S^n) \geq I(Z^{i-1}; Z_i|M, S^n)$
2. $I(Y^{i-1}; Y_i|M, S^n) \geq I(Z^{i-1}; Y_i|M, S^n)$

**Proof:** First of all note that since the channel is memoryless we have

$$(M_1, M_2, M_3, Y_1^{i-1}, Y_2^{i-1}, Y_3^{i-1}, S^{i-1}, S_{i+1}^n) \to (X_i, S_i) \to (Y_{1i}, Y_{2i}, Y_{3i})$$

Just like [4], for any $1 \leq r \leq i-1$

$$I(Z^{r-1}, Y_r^{i-1}; Y_i|M, S^n)$$
$$= I(Z^{r-1}, Y_{r+1}^{i-1}; Y_i|M, S^n)$$
$$\quad + I(Y_r; Y_i|M, S^n, Z^{r-1}, Y_{r+1}^{i-1})$$
$$\geq I(Z^{r-1}, Y_{r+1}^{i-1}; Y_i|M, S^n)$$
$$\quad + I(Z_r; Y_i|M, S^n, Z^{r-1}, Y_{r+1}^{i-1})$$
$$= I(Z^r, Y_{r+1}^{i-1}; Y_i|M, S^n)$$

where the inequality follows from the memorylessness of the channel and the fact that $Y$ is less noisy than $Z$, i.e.

$$I(Y_r; Y_i|M, S^n, Z^{r-1}, Y_{r+1}^{i-1}) \geq I(Z_r; Y_i|M, S^n, Z^{r-1}, Y_{r+1}^{i-1}).$$

Proof of the second part follows the same as the first part with negligible variations. ∎

Now we stick to the proof of the converse

$$nR_3 = H(M_3) = H(M_3|S^n) = H(M_3|S^n, Y_3^n)$$
$$+ I(M_3; Y_3^n|S^n) \leq H(M_3|Y_3^n) + I(M_3; Y_3^n|S^n)$$
$$\leq n\epsilon_{3n} + \sum_{i=1}^{n} I(M_3; Y_{3i}|S^n, Y_3^{i-1}) \leq n\epsilon_{3n}$$

$$+ \sum_{i=1}^{n} I(M_3; Y_{3i} | S^{i-1}, S_i, S_{i+1}^n, Y_3^{i-1}) \leq n\epsilon_{3n}$$

$$+ \sum_{i=1}^{n} I(M_3, S^{i-1}, S_{i+1}^n, Y_3^{i-1}; Y_{3i} | S_i) \leq n\epsilon_{3n}$$

$$+ \sum_{i=1}^{n} I(M_3, S^{i-1}, S_{i+1}^n, Y_2^{i-1}; Y_{3i} | S_i) = n\epsilon_{3n} + \sum_{i=1}^{n} I(U_i; Y_{3i} | S_i)$$

where $U_i \triangleq (M_3, S^{i-1}, S_{i+1}^n, Y_2^{i-1})$ and the last inequality follows from Lemma 1.

$$nR_2 = H(M_2) = H(M_2 | M_3, S^n) = H(M_2 | M_3, S^n, Y_2^n)$$

$$+ I(M_2; Y_2^n | M_3, S^n) \leq H(M_2 | Y_2^n) + I(M_2; Y_2^n | M_3, S^n)$$

$$\leq n\epsilon_{2n} + \sum_{i=1}^{n} I(M_2; Y_{2i} | M_3, S^{i-1}, S_i, S_{i+1}^n, Y_2^{i-1}) = n\epsilon_{2n}$$

$$+ \sum_{i=1}^{n} I(M_2, M_3, S^{i-1}, S_{i+1}^n, Y_2^{i-1}; Y_{2i} | M_3, S^{i-1}, S_{i+1}^n, Y_2^{i-1}, S_i)$$

$$= n\epsilon_{2n} + \sum_{i=1}^{n} I(V_i; Y_{2i} | U_i, S_i),$$

where $V_i \triangleq (M_2, M_3, S^{i-1}, S_{i+1}^n, Y_2^{i-1})$. It is clear that for the given choice of $U_i$ and $V_i$, we have the Markov chain (2) satisfied for the channel is assumed to be memoryless.

$$nR_1 = H(M_1 | S^n, M_2, M_3) = H(M_1 | M_2, M_3, S^n, Y_1^n)$$

$$+ I(M_1; Y_1^n | M_2, M_3, S^n) \leq H(M_1 | Y_1^n) + I(M_1; Y_1^n | M_2, M_3, S^n)$$

$$\leq n\epsilon_{1n} + \sum_{i=1}^{n} I(M_1; Y_{1i} | M_2, M_3, S^{i-1}, S_i, S_{i+1}^n, Y_1^{i-1}) \stackrel{(a)}{\leq} n\epsilon_{1n}$$

$$+ \sum_{i=1}^{n} I(X_i; Y_{1i} | M_2, M_3, S^{i-1}, S_i, S_{i+1}^n, Y_1^{i-1}) = n\epsilon_{1n} +$$

$$+ \sum_{i=1}^{n} I(X_i; Y_{1i} | M_2, M_3, S^{i-1}, S_i, S_{i+1}^n)$$

$$- \sum_{i=1}^{n} I(Y_1^{i-1}; Y_{1i} | M_2, M_3, S^{i-1}, S_i, S_{i+1}^n) \stackrel{(b)}{\leq} n\epsilon_{1n}$$

$$+ \sum_{i=1}^{n} I(X_i; Y_{1i} | M_2, M_3, S^{i-1}, S_i, S_{i+1}^n)$$

$$- \sum_{i=1}^{n} I(Y_2^{i-1}; Y_{1i} | M_2, M_3, S^{i-1}, S_i, S_{i+1}^n) = n\epsilon_{1n}$$

$$+ \sum_{i=1}^{n} I(X_i; Y_{1i} | M_2, M_3, S^{i-1}, S_{i+1}^n, Y_2^{i-1}, S_i) = n\epsilon_{1n}$$

$$+ \sum_{i=1}^{n} I(X_i; Y_{1i} | U_i, S_i),$$

where (a) follows from the memorylessness of the channel and (b) follows from Lemma 1.

Now using the standard time sharing scheme, we can easily conclude that any achievable rate triple for the three-receiver less noisy broadcast channel with side information nan-causally available at the transmitter and at the receivers, must satisfy (22) and the proof is complete. ∎

## V. CONCLUSION

We established two achievable rate regions for two special classes of 3-receiver BCs with side information. We also found the capacity region of 3-receiver less noisy BC when side information is available both at the transmitter and at the receivers.